\documentclass[11pt,onecolumn,twoside]{IEEEtran} 

\usepackage{multirow}
\usepackage{amsfonts}
\usepackage{epsfig}
\usepackage{amsmath}
\usepackage{amssymb}
\usepackage[nolist]{acronym}
\usepackage[english]{babel}
\usepackage{cite}
\usepackage{color}
\usepackage{stfloats}
\usepackage{algorithm}
\usepackage{algorithmic}
\usepackage{subfigure}

\begin{document}

\begin{acronym}
\acro{DASH}{dynamic adaptive streaming over HTTP}  
\acro{CDN}{content delivery network}  
\acro{BP}{belief-propagation}  
\acro{CDF}{cumulative density function}
\acro{PDF}{probability density function}  
\acro{QoE}{quality of experience}   
\acro{ILP}{integer linear programming}   
\acro{CRDSA}{contention resolution diversity slotted ALOHA}
\acro{SA}{slotted ALOHA}
\acro{BS}{base station}
\acro{RD}{rate-distortion}
\acro{MAC}{medium access control}
\acro{IC}{interference cancellation}
\acro{IRSA}{irregular repetition slotted ALOHA}
\acro{EEP}{equal error protection}
\acro{UEP}{unequal error protection}
\acro{SIC}{successive interference cancellation}
\acro{BN}{burst node}
\acro{SN}{slot node} 
\acro{MAB}{multi-arm bandit}
\end{acronym}

\newcommand{\pascal}[1]{[\textcolor{red}{\textit{#1}}]}

\title{IRSA Transmission Optimization via Online~Learning}

\author{
Laura Toni 
and Pascal Frossard 
\thanks{L. Toni is with the Electrical and Electronic Department, University College London (UCL), London WC1E 7JE, U.K. (e-mail:  \texttt{l.toni@ucl.ac.uk}). P. Frossard is with \'Ecole Polytechnique F\'ed\'erale de Lausanne (EPFL), Signal Processing Laboratory - LTS4, CH-1015 Lausanne, Switzerland. (e-mail: \texttt{pascal.frossard@epfl.ch.)} 
}
}%
\maketitle
\thispagestyle{empty}

\begin{abstract}
 In this work, we   propose a new learning framework  for optimising transmission strategies   when irregular repetition slotted ALOHA (IRSA) MAC protocol is considered.
  We cast the online optimisation of the MAC protocol design as a multi-arm bandit problem  that exploits the IRSA structure in the learning framework.   Our learning algorithm quickly learns the optimal transmission strategy, leading to higher rate of successfully received packets with respect to baseline transmission optimizations. 
\end{abstract}

\section{Introduction}
 Random  random \ac{SA}  with    \ac{SIC}   \cite{Casini:J07} has been widely considered as an effective  MAC strategy, for its good performance despite the distributed protocol.
In particular, let us   consider a  system  where $L$ sources send  information to a central \ac{BS}, through an  \ac{IRSA} algorithm \cite{Liva:J11}, which is a SA protocol with SIC. The time axis is discretized in MAC frames, each of those composed  of  $M$ time slots.    Visual sensor networks, in which each source sends periodic data to a central BS, represent one possible scenario\footnote{  Random MAC strategies offer scalability, and adaptability to possibly varying $L$, like in the scenario considered in this paper.}.
 Each source sends $K$ source packets per MAC frame and   each source packet is sent  in multiple replicas   within the MAC frame.    Each replica is transmitted within one time slot and replicas sent from the same source are allocated to different slots, which are uniformly selected at random among the $M$ available slots in a MAC frame. The replication rate $l$ is selected  for each source packet at random according to a transmission probability   $\Lambda(x) = \sum_{l=0}^{l_{max}} \Lambda_{l}x^l $, where $\Lambda_{l}$ is the probability that a source transmits $l$ replicas of a given source packet within the MAC frame, and $l_{max}$ is the maximum replication rate. If a transmission slot is selected by only one source (singleton slot), the corresponding message is correctly received. When multiple sources select the same time slot for transmission, a packet collision happens.  The BS   implements the SIC algorithm to recover the collided messages. Finally, the BS defines the \emph{transmission strategy} characterised by the source rate $K$ and the transmission probability  $\Lambda(x)$  in IRSA\footnote{ The model can also be extended to the case in which each source sends only one source packet ($K=1$) and where the BS defines both  $\Lambda(x)$ and $L$.}.

\begin{figure}[t]
\begin{center}
\subfigure[Time slot representation]{
\includegraphics[width=0.4\linewidth,  draft=false]{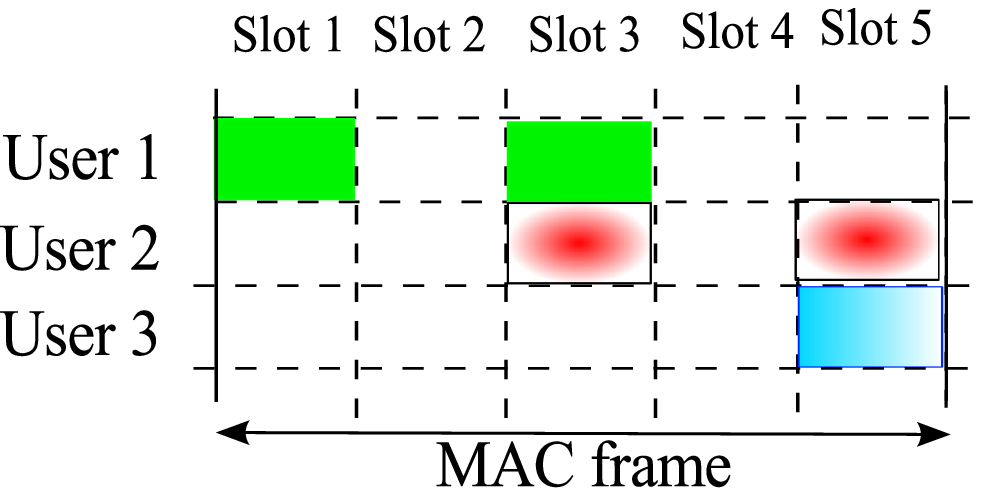}\label{fig:Graph_representation_A}}
\subfigure[Graph-based representation]{
{\includegraphics[width=0.35\linewidth,   draft=false]{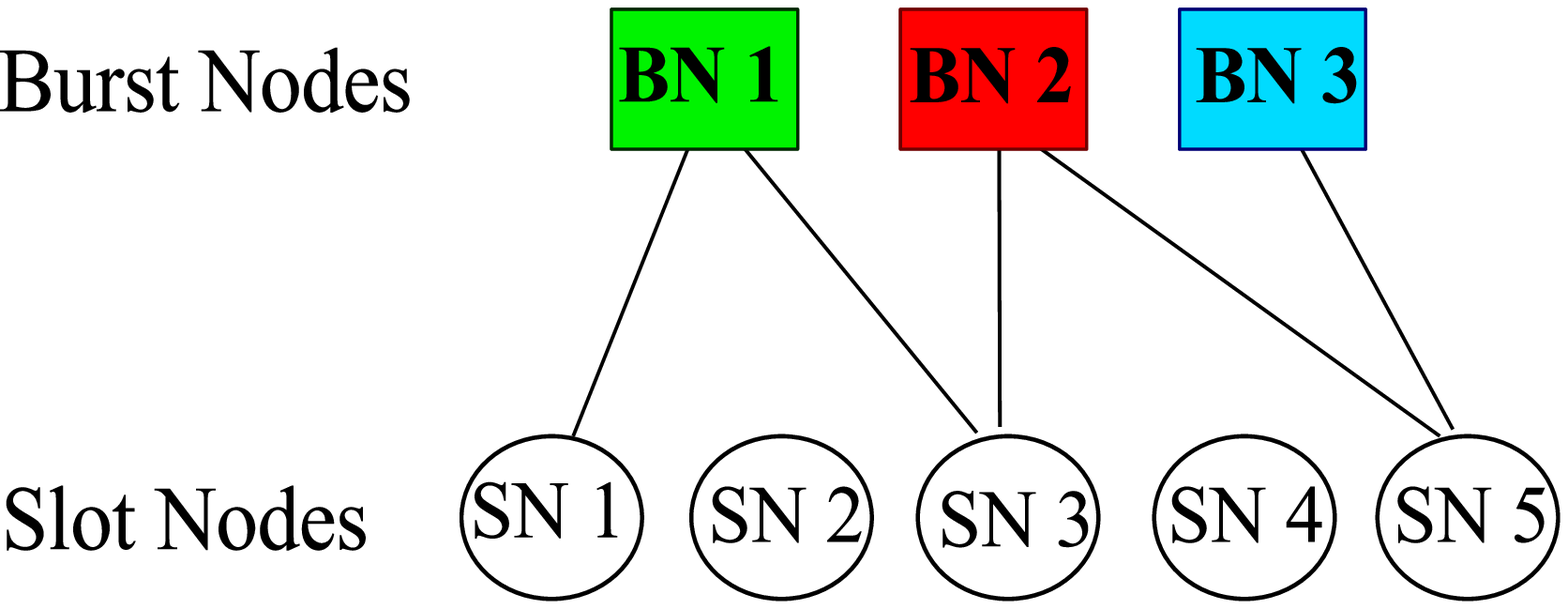}\label{fig:Graph_representation_B}}}
\caption{Example of bipartite graph associated with   IRSA, with $L=3$ sources of one message each ($K=1$) and $M=5$ time slots. Replication rate is $2$   for user $1$ and $2$, and $1$ for user  $3$.}\label{fig:Graph_representation}
\end{center}
\end{figure}

 In the example provided  in   Fig. \ref{fig:Graph_representation_A},   the message sent by user $2$ and user $3$ are not decodable due to collisions in slot $3$ and slot $5$. 
 However, thanks to  \ac{SIC} techniques, the   collision  might be resolved. 
A packet   in a collision-free (singleton) slot is successfully received and decoded, revealing  the slots containing the other replicas.  Their removal via SIC  may turn some of the collided slots into singletons, enabling the recovery of new packets.  This can be seen as message-passing in the bipartite graph Fig, \ref{fig:Graph_representation_B},    opening the possibility of applying  theory of  rateless codes   to IRSA schemes  to analyze both asymptotic  \cite{Liva:J11} and finite-length performance \cite{Paolini:c15,Lazaro:16,Develos:c17}.

In this paper, we are interested in  the {transmission strategy} optimization performed by the BS.   Performance bounds in the asymptotic regime can be easily evaluated and can act as guidelines for a  proper selection of 
the transmission strategy. The accuracy of such guidelines however reduces in non-asymptotic settings.  Exact performance analysis has been recently derived for the finite block length case; unfortunately it is    computationally too expensive for medium block sizes in practical implementations. 
 Frameless ALOHA protocols can compensate for inaccurate performance evaluation, since the MAC frame is not set a priori. New slots are added until a sufficiently high fraction of packets has been decoded. However, this comes at the price of having changing MAC frame size over time.  
 Rather than analytically deriving the IRSA system performance,     we propose to estimate directly the optimal design of IRSA algorithms by online learning.    
 We cast the network optimisation problem as a \ac{MAB} problem \cite{MAB-Tutorial}, i.e., sequential decision strategy in which a decision maker needs to select the optimal ``arm". The \ac{BS} is the \emph{decision maker} who  measures the overall performance (\emph{reward}) of each arm (transmission strategy) and learns the optimal network design (\emph{optimal actions}) by trial and error.  To improve the learning performance, we design a specific IRSA-based MAB algorithm where we infer the IRSA structure into the learning problem in order to improve on the low convergence of classical MAB solutions. 
Simulation results show that our learning algorithm reaches higher mean reward   with respect to  transmission strategies designed based on $i)$ classical bandit problems  $ii)$ analytical performance studies.

%
%



\section{Online IRSA optimization}
\label{sec:algo} 
\subsection{Problem Formulation}
The performance of the system is   measured by an utility function $U(r)$ that is a non-decreasing function of the number of per-source decoded packets $r\leq K$ at the BS. In this work, we consider a generic utility function of the form 
\begin{align}\label{eq:utility}
U(r)  = w \log(r+1)
\end{align}
where $w$ is a scaling factor.  The above utility function is typical in multimedia transmission, (e.g., visual sensor networks) and $w$ reflects different priorities among different sources. However, the proposed learning strategy applies to any other increasing utility function. 

In the above framework, optimising the system performance consists in finding the best transmission strategy (i.e.,   $\Lambda(x)^{\star}$,  $K^{\star}$), such that the overall mean utility function per source is maximised. Formally, the optimisation problem is 
\begin{align}\label{eq:probl1}
 (\Lambda(x)^{\star}, K^{\star}):  \arg\max_{\{\Lambda(x),K\}} \ &  \sum_{r=0}^K w \log(r+1) P(r| \Lambda, K ; M,L) \nonumber \\
\text{s.t. } &L K   \leq M
\end{align}
where $P(r| \Lambda, K ; M,N)$ is the probability of decoding $r$ packets in the scenario with $L$ sources transmitting $K$ packets over $M$ time slots. The network traffic $L K/M$ is finally constrained to be lower than $1$, which corresponds to the stability limit of the SA MAC protocol~\cite{Liva:J11}.

\subsection{MAB problem formulation}

We now cast the above optimisation as a  \ac{MAB} problem. 
 In our case, the BS is the decision maker, which can periodically adjust the transmission strategy. Let $t$ denote  the decision opportunity timestamp\footnote{ We consider a decision every MAC frame but less frequent decisions can also be considered.}. Between two consecutive decision opportunities, the BS observes the overall system performance (i.e., the number of  correctly decoded packets) that is the instantaneous reward. Based on this observation, it selects the next action, which is then communicated to all sources.   The overall goal is to  minimize the experienced regret, i.e., the loss due to the fact that the globally optimal policy is not achieved all the times. 

Let $a\in \mathcal{A}$ be one possible transmission strategy, with $\mathcal{A}$ being the set of all possible strategies (possible arms). In our IRSA problem, an arm $a$ is associated to the transmission strategy defined by $\{\Lambda_{a}(x)=[\Lambda_{1}^{(a)}, \Lambda_{2}^{(a)}, \ldots, \Lambda_{l_{\text{max}}}^{(a)}], K_{a}\}$. Let further $\mu_a=\sum_{r=0}^K w \log(r+1) P(r| \Lambda_a(x), K_a ; M,L) $ be the mean reward of the arm $a$.    At time $t$, the learner selects the action $a$ and experiences an instant payoff $X_{a,t}$, which is a realization of a random variable with unknown distribution  $F_a$ and  unknown mean value $\mu_a$. 
 When the performance value is known, $P(r| \Lambda_a(x), K_a ; M,L)$ could be derived and (2) could be solved numerically. Conversely, if there is no precise value (or  no  low-complexity evaluation) of the performance evaluation,   we can learn $P(r| \Lambda_a(x), K_a ; M,L)$ from experience, i.e., by trial-and-error.  
 During the learning process  suboptimal actions might be selected, leading to a cumulative regret after $t$ decisions that is  
$R(t)= t \mu^{\star} - \sum_{i=1}^t X_{\alpha_i, i}$
 with $\alpha_i$  being the arm selected at the $i^{th}$ decision opportunity and $\mu^{\star}$ the mean reward of the optimal arm. 
 \begin{algorithm}[!t] 
  \caption{UCB}
  \begin{algorithmic}
\STATE     {\bf Input:} $n$ (horizon), $A$ (number of arms)  
\STATE     {\bf Initialize:} set $\hat{\mu}_{a,0}$   from the asymptotic theoretical analysis.
\STATE    $t=1$
\WHILE {$t\leq n$} 
        \STATE $\alpha_t = \arg\max_{a\in\mathcal{A}}    \left\{\hat{\mu}_{a,t} + \beta \sqrt{\frac{2 \log t }{N_{a,t}}} \right\}$ 
        \STATE Select arm   $\alpha_t$, receive reward $X_{{\alpha_t},t}$
        \STATE   $\hat{\mu}_{\alpha_t,t} = (N_{\alpha_t} \hat{\mu}_{\alpha_t,t} + X_{{\alpha_t},t})/(N_{\alpha_t}+1)$
        \STATE  $N_{\alpha_t}=N_{\alpha_t}+1$
        \STATE  $t=t+1$
\ENDWHILE 
\end{algorithmic}
\end{algorithm}

 The classical algorithm to solve MAB problems is the upper confidence bound (UCB) algorithm~\cite{MAB-Tutorial} (see Alg.~1). The UCB solves MABs using the optimism in face of uncertainty principles in order to find the best tradeoff between exploration and exploitation.  Rather than selecting the action with the highest estimated reward $\hat{\mu}_{a,t}$, the UCB selects the reward with the highest bandit index    $ b_{a,t} = \hat{\mu}_{a,t} + \beta \sqrt{ {2 \log t } /  {N_{a,t}}}$,   
with $N_{a,t}$ being the number of times the arm $a$ has been selected up to $t$, and $\beta$ a multiplicative factor.  The  second term in the bandit index   represents   an upper confidence bound  that reflects the uncertainty on  the estimates of $\hat{\mu}_{a,t}$   
   Asymptotically, the  UCB algorithm  minimizes the regret  $R(t)$ and  therefore it maximizes the mean reward  $\sum_{i=1}^t X_{\alpha_i, i}$. 
Let define $X_{a,t}= (1/L) \sum_{l=1}^L w\log(r_{a,t}^{(l)}+1)$, with $r_{a,t}^{(k)}$ being the number of decoded packets for source $k$ with action $a$ taken at time $t$. This means that asymptotically the  UCB optimizes the expected utility function  for each source, i.e.,  $\sum_{r=0}^K w \log(r+1) P(r| \Lambda, K ; M,L)$. Defining the set of actions as $\mathcal{A}:\{a=(\Lambda_a(x),K_a) | L K_a \leq M\}$ the   UCB algorithm asymptotically optimises the    problem in \eqref{eq:probl1}.

\subsection{Proposed online learning algorithm}

Classical UCB methods can be improved if prior information is inferred in the learning process. This is the reason why we propose an online learning solution based on a Bayesian UCB (or Bayes-UCB) algorithm  \cite{kaufmann2012bayesian} that allows us to infer information about the reward distribution. We represent the uncertainty about the system in terms of variance of the process rather than a confidence bound of the estimate as in classical UCB methods. Thus, the learning process consists in selecting at each time instant $t$ the arm $\alpha_t$ that maximises the following function 
$\alpha_t =  \arg\max_{a}  \{ \hat{\mu}_{a,t} + \beta  \sigma_{a,t}\}$,  
 with  $\hat{\mu}_{a,t}$ and $\sigma_{a,t}^2$ being respectively the  mean and variance of the reward of action $a$ estimated at the decision opportunity $t$, and $\beta$ being   a multiplicative factor. 
 The Bayes-UCB algorithm thus automatically builds confidence intervals based on a Kullback-Leibler divergence, which best fits the geometry of the problem (see Appendix).  
 
The Bayes-UCB algorithm initiates a mean and variance per arm.  As there exists no \emph{exact} information about the mean and the variance of the process in our IRSA system, we derive estimates from the \emph{asymptotic} setting~\cite{Liva:J11},   and  under the assumption  that each source independently assigns packets to transmission slots. In  practice, the independency holds only among packets sent from different sources, therefore   less collisions are actually experienced than the number predicted from the theory.  The derived estimate is then refined by the Bayes-UCB algorithm at each decision opportunity.   
 
 \begin{algorithm}[!t]
  \caption{Bayesian UCB}\label{algo:Bayesian-UCB}
  \begin{algorithmic}
\STATE     {\bf Input:} $n$ (horizon), $A$ (number of arms) 
\STATE     {\bf Initialize:} set $\hat{\mu}_{a,0}$ and $\sigma^2_{a,0}$ 
\STATE    $t=1$
\WHILE {$t\leq n$} 
	\STATE  
		$\alpha_t =  \arg\max_{a}   \{ \hat{\mu}_{a,t} + \beta  \sigma_{a,t}\}  $ 
        \STATE Select arm $\alpha_t$, receive reward $X_{{\alpha_t},t}$
        \STATE   Perform Bayesian update to derive $\hat{\mu}_a(t)$, and $\sigma_a(t)$

        \STATE  $t=t+1$
\ENDWHILE 
\end{algorithmic}
\end{algorithm}

   For the arm $a$ with transmission strategy $( \Lambda_a(x), K_a)$, we evaluate $P_e(\Lambda_a)$ as the probability for a packet to be lost after   SIC.  This is derived from the asymptotic analysis  in   \cite{Liva:J11}, exploiting the     SIC convergence analysis  and the probability of a burst node  edge being not resolved at an iteration, \cite[Eq~2]{Liva:J11}.   Due to  the independency between sources, 
   the probability for one source to correctly decode $r$ source packets is 
\begin{align}\label{eq:PLR}
 P(r| \Lambda_a, K_a ; M,L) = {{K_a}\choose{r} } P_e(\Lambda_a)^{r}[1-P_e(\Lambda_a)]^{K_a-r} .
 \end{align}
We then note that the mean reward of the utility given in \eqref{eq:utility} is given by the logarithm of a binomial random variable. Therefore, by expanding the logarithm operator as a Taylor series in $x=K_a  P_e(\Lambda_a)$ and computing the binomial sum term by term, the mean reward and the reward variance of action $a$ can be estimated with 
\begin{align}\label{eq:mu-a}
\mu_{a} = \log\left(K_a P_e(\Lambda_a) +1 \right), \ \ 
\sigma^2_{a} = \frac{ K_a  P_e(\Lambda_a) (1- P_e(\Lambda_a)) }{( P_e(\Lambda_a) K_a+1)^2 } \nonumber
\end{align}
The initial estimates $\hat{\mu}_{a,0}$ and $\sigma^2_{a,0}$ are thus derived from the above equations, and the learning algorithms proceeds as described in   Alg. \ref{algo:Bayesian-UCB}.
 
\subsection{  Computational Complexity }
Our proposed learning strategy is a low-complexity algorithm compared to transmission strategies designed based on finite-length performance analysis.  At each decision opportunity, the only operations that need to be computed are $i)$ the selection of arms (which is a maximization of a known vector), $ii)$ the update of the mean reward only for the selected arm (a weighted sum).  The initialization process requires the evaluation of 
\eqref{eq:PLR} for all possible arms. The cardinality of the arm set depends on $l_{max}$ and the maximum value of $K$, but not on the MAC frame, neither on the number of sources. Moreover, this initialization step can be further simplified by approximating the packet error probability by a binary condition given by \cite[Eq~7]{Liva:J11}. 

The exact finite length analysis \cite{Lazaro:16,Develos:c17} has in contrary a complexity that scales exponentially.   The combinatorial approach in \cite{Develos:c17} spans the all possible combinations of transmission realizations with  a complexity $\mathcal{O}(|\mathcal{N}| \, l_{max}^{LK})$, with  
 $|\mathcal{N}|$ being the cardinality of the set $\mathcal{N}$, which is the set of possible edge-realizations in the bipartite graph. This is given by    $|\mathcal{N}| \leq   \binom{\hat{\mathcal{C}}+M-1}{t}$, with $\hat{\mathcal{C}}=\sum_{n=0}^{LK-2} \binom{LK}{M}$.  A computational complexity that scales exponentially with the MAC frames and number of users is achieved also in \cite{Lazaro:16}, where the system performance evaluation is derived based on a finite Markov state machine and each state is characterized by the size of the cloud and ripples (with different degrees).

\section{Simulation results} 
\label{sec:results} 

\begin{figure}[t]
\begin{center}
\includegraphics[width=.45\linewidth,  draft=false]{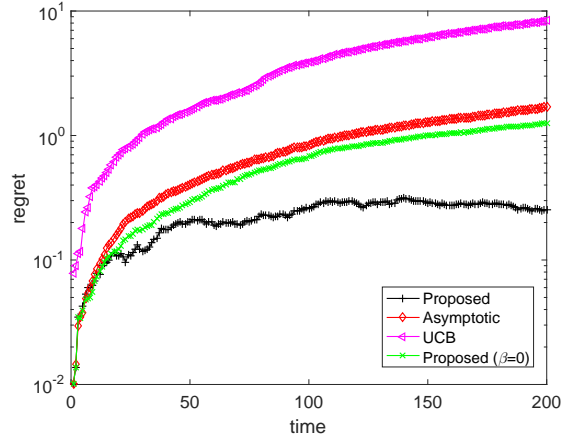}
\caption{Cumulative regret vs. time  (in terms of decision opportunity) for the scenario with $L=20$   and $M=300$. The transmission strategy is $\Lambda(x)=0.75x^2+0.25x^3$, while $K$ is optimized.  In this settings, the optimal $K$ is $K=7$ while the asymptotic baseline method sets the best $K$ to $5$.}\label{fig:normalized_1_N20_M300_Lambda14}
\end{center}
\end{figure}

We now provide simulation results that show the performance of the proposed learning strategies with respect to $(i)$ classical UCB algorithm, and $(ii)$ MAC protocol optimisation based on asymptotic strategies.  Results are provided in terms of both regret ($R(t)= t \mu^{\star} - \sum_{i=1}^t X_{\alpha_i, i}$) and experienced reward.  Note that the finite-length analysis   leads to the optimal decision strategy at each time step, meaning a null cumulative regret. Therefore, the regret shows the tradeoff between complexity (in evaluating the finite length analysis) and performance approximation   (due to the learning process).   
The proposed algorithm (labeled ``Proposed'') is compared with the classical UCB algorithm, both with $\beta=1$. We also compare to a baseline solution (denoted ``Asymptotic") where the optimal transmission strategy is not learned but rather computed by assuming an asymptotic behavior of the system  (i.e., very large MAC frame duration). Namely, $P_e(\Lambda_a)$ in \eqref{eq:PLR} is evaluated from \cite[Eq~2]{Liva:J11} and  the optimisation problem in \eqref{eq:probl1} is   solved numerically.  
Finally, we also consider a learning algorithm (labeled as ``Proposed ($\beta=0$)") that does not take into account the confidence bound of the estimation. Namely, we consider our proposed algorithm with $\beta=0$, which actually corresponds to the UCB algorithm with $\beta=0$.

We first consider the case in which the transmission probability $\Lambda(x)$ is fixed and the actions correspond to different number of packets per source $K$. Fig. \ref{fig:normalized_1_N20_M300_Lambda14} provides the cumulative regret for the scenario with $300$ time slots per MAC frame, $20$ sources and  $\Lambda(x)=0.75x^2+0.25x^3$. The proposed learning solution (IRSA Bayes-UCB) generally outperforms all the other learning strategies. Finally, in this specific setting, the optimisation based on the asymptotic theoretical performance achieves good performance, even if it performs worse than our proposed learning strategy. Its performance however degrades in other settings as shown next.  

%

\begin{figure}[!htb]
  \centering
    \begin{minipage}{.45\textwidth}
       \centering
        \includegraphics[width=0.9\linewidth,  draft=false]{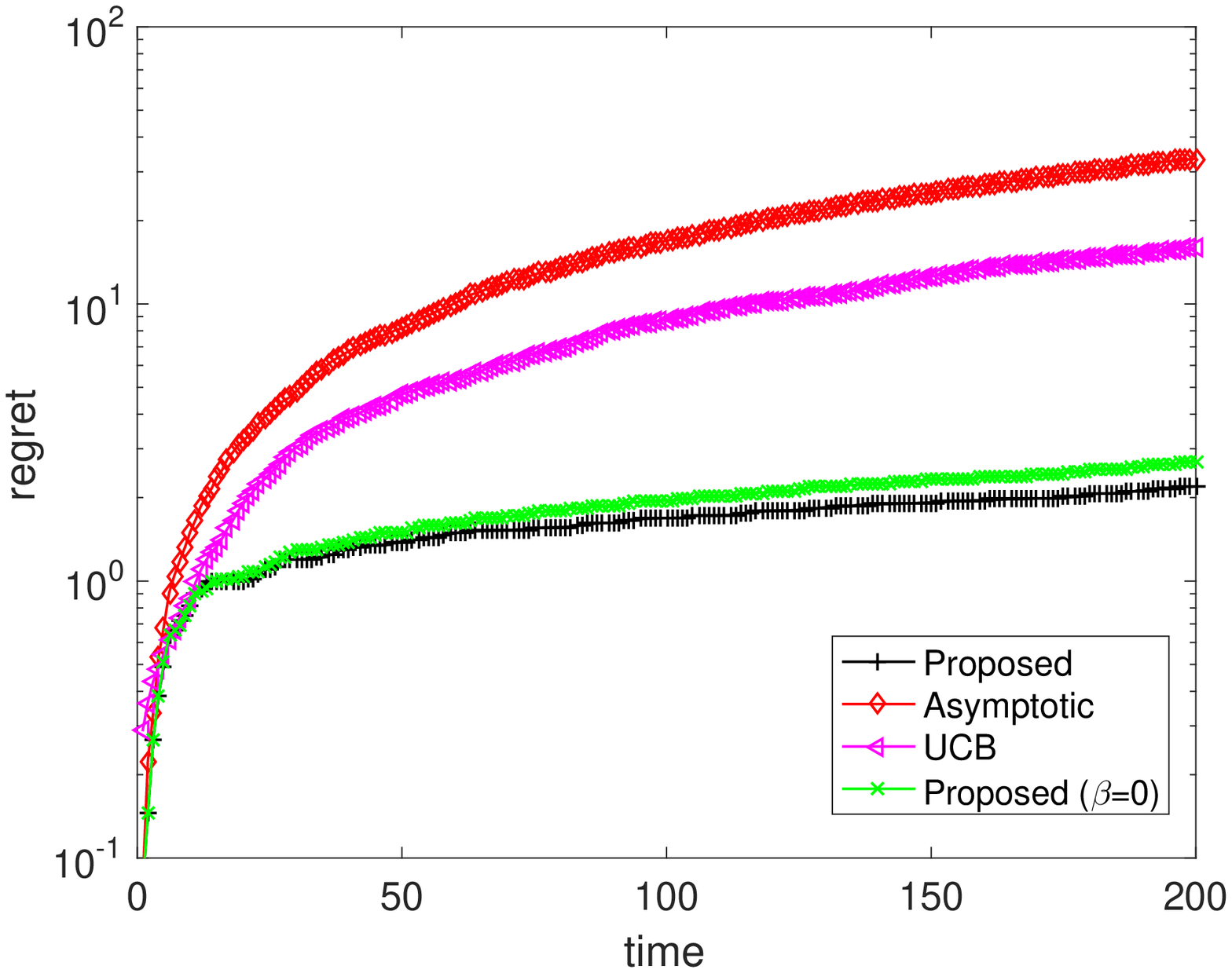}
  	\caption{Cumulative regret vs. time  (in terms of decision opportunity) when  jointly optimizing   $K$ and $\Lambda(x)$ for the scenario of $L=20$ and $M=300$. }\label{fig:normalized_1_N20_M300_Lambda1_Joint}
     \end{minipage}%
     \hspace{0.5cm}
    \begin{minipage}{0.45\textwidth}
      \centering
        \includegraphics[width=0.9\linewidth,  draft=false]{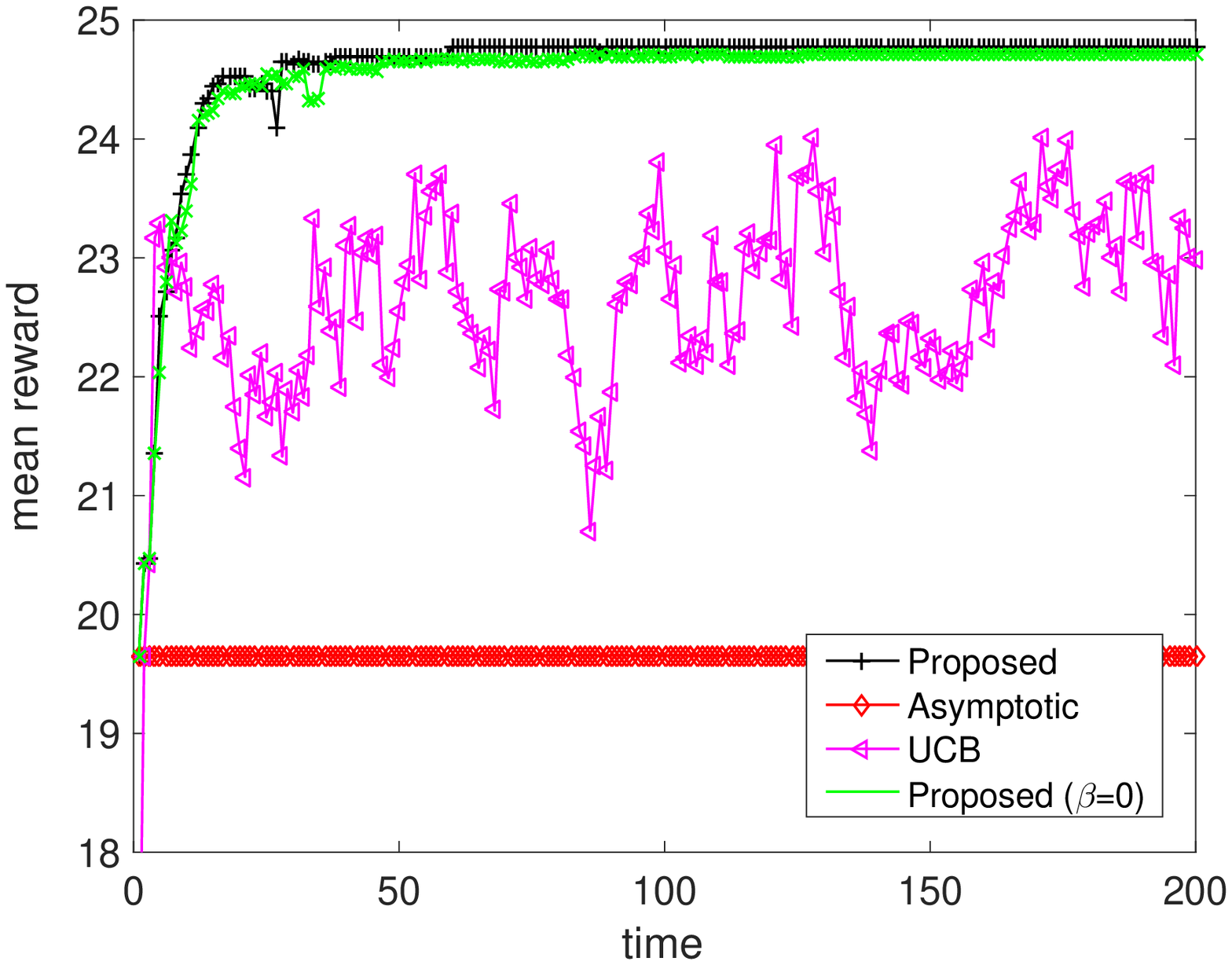}
      \caption{Mean reward experienced over time  (in terms of decision opportunity)  when  jointly optimizing   $K$ and $\Lambda(x)$ when $L=20$ and $M=300$.}\label{fig:normalized_mu_1_N20_M300_Lambda1_Joint}
    \end{minipage}
\end{figure}

In Fig. \ref{fig:normalized_1_N20_M300_Lambda1_Joint}, the cumulative regret is provided when  $\Lambda(x)$ and $K$ are jointly optimised in the scenario of $20$ sources and $300$ time slots.   The candidates $\Lambda(x)$  are defined as  $\Lambda(x)=a_1x^2+a_2x^3+a_3x^8$, with the coefficients $a_i$ ranging from $0$ to $1$, i.e., $a_i\in[0$:$0.25$:$1], i=1,2,3$. The candidate values of $K$ range from $1$ to $\lfloor M/L\rfloor$. 
Moreover, note that the gain achieved by the proposed method (IRSA Bayes-UCB) with respect to the UCB  algorithm has increased compared to the previous setting. This is due to the fact that the action space in this scenario has increased, leading to a slower learning curve and therefore to a higher gain in exploiting prior information while learning, as implemented in our algorithm.   It is worth mentioning that the ``Proposed ($\beta=0$)" baseline performs very closely to the proposed learning. This is due to the small randomness of the reward for all arms beyond the traffic threshold value $G^{\star}$, as explained in the Appendix. However, the randomness in the remaining arms makes the ``Proposed ($\beta=0$)" worse than our IRSA Bayes-UCB algorithm. 
Finally, we provide the mean reward experienced over time in Fig. \ref{fig:normalized_mu_1_N20_M300_Lambda1_Joint}. The reward per instant (averaged over $100$ runs) confirms the gain of the proposed learning strategy with respect to the baseline methods. 

\section{Conclusions}

We have proposed a learning framework for designing optimised IRSA transmission strategies. We have casted the optimal resource allocation and transmission rate optimisation as a multi-arm bandit (MAB) problem. We have then implemented a specific MAB algorithm that is able to exploit the initial knowledge about IRSA schemes in the form of asymptotic theoretical performance. This allows us to infer the structure of IRSA in MAB problems and improve the learning efficiency of the algorithm. Simulation results have validated our theoretical analysis and demonstrated the gain of the proposed learning strategy in all tested settings.
 
 \
 
 \
 
\section*{Appendix \\ Suboptimality of UCB}  The key intuition of the UCB algorithm is the following. At time $t$, the estimated payoff for arm $a$ is $\hat{\mu}_{a,t}$ and it   differs from the estimated one of at most $U$  (i.e., $ |\hat{\mu}_{a,t} - \ {\mu}_{a}| \leq U$) with probability $p$  given by~\cite{MAB-Tutorial}
$ p \leq e^{-2N_{a,t} U^2}.$
Imposing a probability $p$ decreasing with time (e.g., $p=t^{-4}$) 
leads to $U_{\text{max}}= \sqrt{2\log t/ N_{t,a}}$. To ensure  that the learner selects the optimal action as $t\rightarrow \infty$, the estimated  reward $\hat{\mu}_{a,t}$ is added to the confidence bound $U_{\text{max}}$, which leads to Algorithm 1 in Sec. \ref{sec:algo}.  However, arms with a traffic lower than  the threshold $G^{\star}$ will experience an almost deterministic payoff. 
Denoting the network traffic as  $G=L K/M$,   for $G<G^{\star}$ packets are received with probability almost $1$, while for $G>G^{\star}$ the probability of correctly receiving the packets   collapses to $0$. This   waterfall effect  is typical for IRSA  \cite{Liva:J11}.  This means that   $ |\hat{\mu}_{a,t} - {\mu}_{a}| \leq \epsilon$ with probability almost $1$, for small $\epsilon$ and $t$. Therefore, the confidence bound actually differs from different arms. For this reason, the UCB results in a suboptimal algorithm. Conversely, the Bayesian method allows us to infer this heterogeneity in the uncertainty of the arms, by imposing different variance values or the reward for different arms.

\bibliographystyle{IEEEtran}
\bibliography{MAC_Bandit,LDPC_SlottedAloha}

\end{document}